\documentclass[%
aip,
 amsmath,amssymb,
 reprint,%
]{revtex4-1}

\usepackage{graphicx}
\usepackage{dcolumn}
\usepackage{soul}
\usepackage{xcolor}
\sethlcolor{yellow}
\usepackage{bm}
\usepackage[utf8]{inputenc}
\usepackage[T1]{fontenc}
\usepackage{mathptmx}
\usepackage{etoolbox}
\usepackage{tabularx}
\usepackage{caption} 
\usepackage{titlesec}
\titlespacing{\section}{10pt}{10pt}{0pt}

\makeatletter
\def\@email#1#2{%
 \endgroup
 \patchcmd{\titleblock@produce}
  {\frontmatter@RRAPformat}
  {\frontmatter@RRAPformat{\produce@RRAP{*#1\href{mailto:#2}{#2}}}\frontmatter@RRAPformat}
  {}{}
}%
\makeatother
\usepackage{titlesec}

\begin{document}

\preprint{AIP/123-QED}

\title[Low cross-plane thermal conductivity of $\kappa$-Ga$_2$O$_3$ thin films] {Low cross-plane thermal conductivity of $\kappa$-Ga$_2$O$_3$ thin films}
\author{Seungwon Park}
\affiliation{ 
Department of Mechanical Engineering, Carnegie Mellon University, Pittsburgh, Pennsylvania 15213, USA
}%
\author{Jingyu Tang} 
\affiliation{%
Department of Materials Science and Engineering, Carnegie Mellon University, Pittsburgh, Pennsylvania 15213, USA
}%
\author{Minsung An}
\affiliation{ 
Department of Mechanical Engineering, Carnegie Mellon University, Pittsburgh, Pennsylvania 15213, USA
}%
\author{Alan J. H. McGaughey}
\affiliation{ 
Department of Mechanical Engineering, Carnegie Mellon University, Pittsburgh, Pennsylvania 15213, USA
}%
\author{Lisa M. Porter}
\affiliation{%
Department of Materials Science and Engineering, Carnegie Mellon University, Pittsburgh, Pennsylvania 15213, USA
}%

\author{Robert F. Davis}
\affiliation{%
Department of Materials Science and Engineering, Carnegie Mellon University, Pittsburgh, Pennsylvania 15213, USA
}%

\author{Jonathan A. Malen}
\email{jonmalen@andrew.cmu.edu}
\affiliation{ 
Department of Mechanical Engineering, Carnegie Mellon University, Pittsburgh, Pennsylvania 15213, USA
}%
\affiliation{%
Department of Materials Science and Engineering, Carnegie Mellon University, Pittsburgh, Pennsylvania 15213, USA
}%

\date{\today}
\begin{abstract}
The thermal conductivities of (001) $\kappa$-Ga$_2$O$_3$ thin films deposited on $c$-plane sapphire substrates were measured with frequency-domain thermoreflectance (FDTR). The thermal conductivity increases from {\color{black}1.83 $^{+0.19}_{-0.18}$ W/m$\cdot$K to 2.26 $^{+0.35}_{-0.40}$ W/m$\cdot$K} as the thicknesses increase from {\color{black}159 nm to 689 nm.} {\color{black}This weak thickness dependence indicates that the majority of the intrinsic phonon mean free paths are smaller than the range of investigated film thicknesses. First-principles-based calculations that include phonon-phonon and phonon-boundary scattering capture this size-dependent behavior and support our experimental observation.} Temperature-dependent measurements {\color{black} of a 689 nm film} from 293 to 400 K show a decrease in thermal conductivity with increasing temperature due to enhanced phonon-phonon scattering.  A weaker temperature dependence is observed in {\color{black}a 159 nm film} where phonon-boundary scattering becomes more prominent. {\color{black}At comparable film thicknesses, our measured $\kappa$-Ga$_2$O$_3$ films exhibit lower values of cross-plane thermal conductivity than the other experimentally measured Ga$_2$O$_3$ polymorphs.} Our experimental observations are supported by prior predictions, which suggest that the low thermal conductivity of $\kappa$-Ga$_2$O$_3$ results from its complex crystal structure, which leads to increased anharmonicity and lower phonon group velocities relative to other polymorphs. Thermal management strategies, such as integration with high-thermal-conductivity substrates, may thus be required to mitigate self-heating for $\kappa$-Ga$_2$O$_3$-based electronic devices.
\end{abstract}
\maketitle
Gallium oxide (Ga$_2$O$_3$) is an ultra-wide bandgap (UWBG) semiconductor that occurs in four polymorphs, $\alpha$-, $\beta$-, $\gamma$-, and $\kappa$-. \cite{roy_polymorphism_1952} $\kappa$-Ga$_2$O$_3$ is the only polymorph lacking inversion symmetry, resulting in spontaneous polarization and piezoelectric behavior. \cite{tang_thermal_2024, he_high_2023, maccioni_phase_2016} With a piezoelectric strain constant $d_{33}$ that is twice that of AlN, $\kappa$-Ga$_2$O$_3$ has the potential to be an emerging piezoelectric semiconductor for radio frequency (RF) resonators in high electron mobility transistors (HEMTs).\cite{chen__2022} Its spontaneous polarization also leads to a high density two-dimensional electron gas (2DEG) at the $\kappa$-Ga$_2$O$_3$/AlN interface without doping.\cite{wang_-ga2o3_2020} $\kappa$-Ga$_2$O$_3$  can also be stacked with $\beta$-Ga$_2$O$_3$ to create a $\kappa$-Ga$_2$O$_3$/$\beta$-Ga$_2$O$_3$ heterojunction, which can be exploited for devices that rely on interfacial depletion fields to enable efficient carrier separation.\cite{lu_ga2_2025} 
{\color{black}Although $\kappa$-Ga$_2$O$_3$ is thermodynamically metastable,\cite{tang_thermal_2024, maccioni_phase_2016} it exhibits substantial kinetic thermal stability and typically transforms to the $\beta$-phase only after annealing at approximately 700-1000 °C, enabling a large window for device fabrication and high-temperature operation.\cite{lee_high_2021, tang_thermal_2024, tang_kinetic_2026}}
$\beta$-Ga$_2$O$_3$ has attracted significant interest for applications in next-generation power electronics and deep-ultraviolet photodetectors due to its ultrawide bandgap ($\sim$4.8 eV) and its thermal stability.\cite{xue_overview_2018, higashiwaki_-ga2o3_2022} $\kappa$-Ga$_2$O$_3$, with a bandgap ($\sim$4.9 eV) \cite{oshima_epitaxial_2015} similar to that of $\beta$-Ga$_2$O$_3$, has the potential to be used in similar applications. However, $\beta$-Ga$_2$O$_3$'s relatively low thermal conductivity (11-27 W/m·K)\cite{guo_anisotropic_2015} compared to other wide bandgap semiconductors, such as GaN ($\sim$210 W/m·K)\cite{jezowski_thermal_2003} and AlN ($\sim$320 W/m·K),\cite{koh_bulk-like_2020} poses a limitation to effective thermal management in RF and power devices.

The thermal conductivities of different Ga$_2$O$_3$ polymorphs have been reported in the literature. Szwejkowski \textit{et al.} showed size effects in the thermal conductivity of  $\beta$-Ga$_2$O$_3$ thin film, as the thermal conductivity increases from 0.3 W/m·K to 9.0 W/m·K with thickness increasing from 12 nm to 600 nm.\cite{szwejkowski_size_2015}  Song \textit{et al.} also reported the increase in the thermal conductivity of $\beta$-Ga$_2$O$_3$ thin films from 6.3  $\pm$ 0.6 W/m·K to 10.8  $\pm$ 1.2 W/m·K with thickness increasing from 164 nm to 2500 nm. \cite{song_thermal_2021} Our previous work has demonstrated a size dependence of $\gamma$-Ga$_2$O$_3$ thin films, which increase from $2.3_{-0.5}^{+0.9}$ W/m·K to $3.5\pm 0.7$ W/m·K for films with thicknesses of 75 nm to 404 nm.\cite{park_size_2025} These size effects in the thermal conductivity of Ga$_2$O$_3$ occur because boundary scattering reduces contributions from phonons with intrinsically long mean free paths (MFPs).\cite{chen_thermal_1998, cahill_nanoscale_2014} Karim \textit{et al.} reports the thermal conductivity of 1.2 $\mu$m thick $\alpha$-Ga$_2$O$_3$ to be 11.9  $\pm$ 1.0 W/m·K,\cite{karim_thermal_2023} which is the highest of all polymorphs at comparable thicknesses. Several computational studies have reported the lattice thermal conductivity of $\kappa$-Ga$_2$O$_3$ along the (001) direction, which ranges from 1 to 4 W/m$\cdot$K.\cite{yang_lattice_2024, xiao_insights_2024, sun_neuroevolution_2023, sun_influence_2024} Xiao \textit{et al.}\cite{xiao_phasedependent_2024} have measured the thermal conductivity of 50, 65, and 89 nm thick (001) $\kappa$-Ga$_2$O$_3$ to be $\sim$1-3 W/m$\cdot$K, with a negligible size dependence. However, extensive measurements of the thermal conductivity of $\kappa$-Ga$_2$O$_3$ over a wider thickness range have not been reported in the literature. 

In this paper, we report the thermal conductivity of $\kappa$-Ga$_2$O$_3$ thin films deposited on $c$-plane sapphire substrates, where the measurements were taken along the (001) direction of the films and over film thicknesses ranging from {\color{black}159 nm to 689 nm. To support our experimental results, we also employ first-principles-based calculations of the size-dependent thermal conductivity of $\kappa$-Ga$_2$O$_3$. In addition, we} report the temperature-dependent thermal conductivity of {\color{black}159 nm and 689 nm} $\kappa$-Ga$_2$O$_3$ films at temperatures ranging from 293 K to 400 K. The results are compared with experimentally measured thermal conductivities of other Ga$_2$O$_3$ polymorphs reported previously. 


Six $\kappa$-Ga\textsubscript{2}O\textsubscript{3} films with thicknesses ranging from {\color{black}159 nm to 689 nm} were grown on (0001) Al\textsubscript{2}O\textsubscript{3} ($c$-sapphire) substrates at 530 $^{\circ}$C using a vertical, low-pressure, cold-wall metal-organic chemical vapor deposition reactor. Triethylgallium (TEGa) and ultra-high purity O\textsubscript{2} (5N) were used as the precursors, with N\textsubscript{2} serving as the carrier gas for the TEGa. The TEGa bubbler was maintained at 20 $^{\circ}$C with 45 sccm N\textsubscript{2} carrier gas flowing through the bubbler, resulting in a TEGa molar flow rate of 70 $\mu$mol/min. The O\textsubscript{2} flow rate was fixed at 500 sccm, corresponding to an O/Ga (VI/III) ratio of 635. The detailed growth parameters and comprehensive film characterization have been reported in our previous publications.\cite{jiang_layered_2022,tang_thermal_2024}

\begin{figure}[h]
    \centering
    \includegraphics[width=\linewidth]{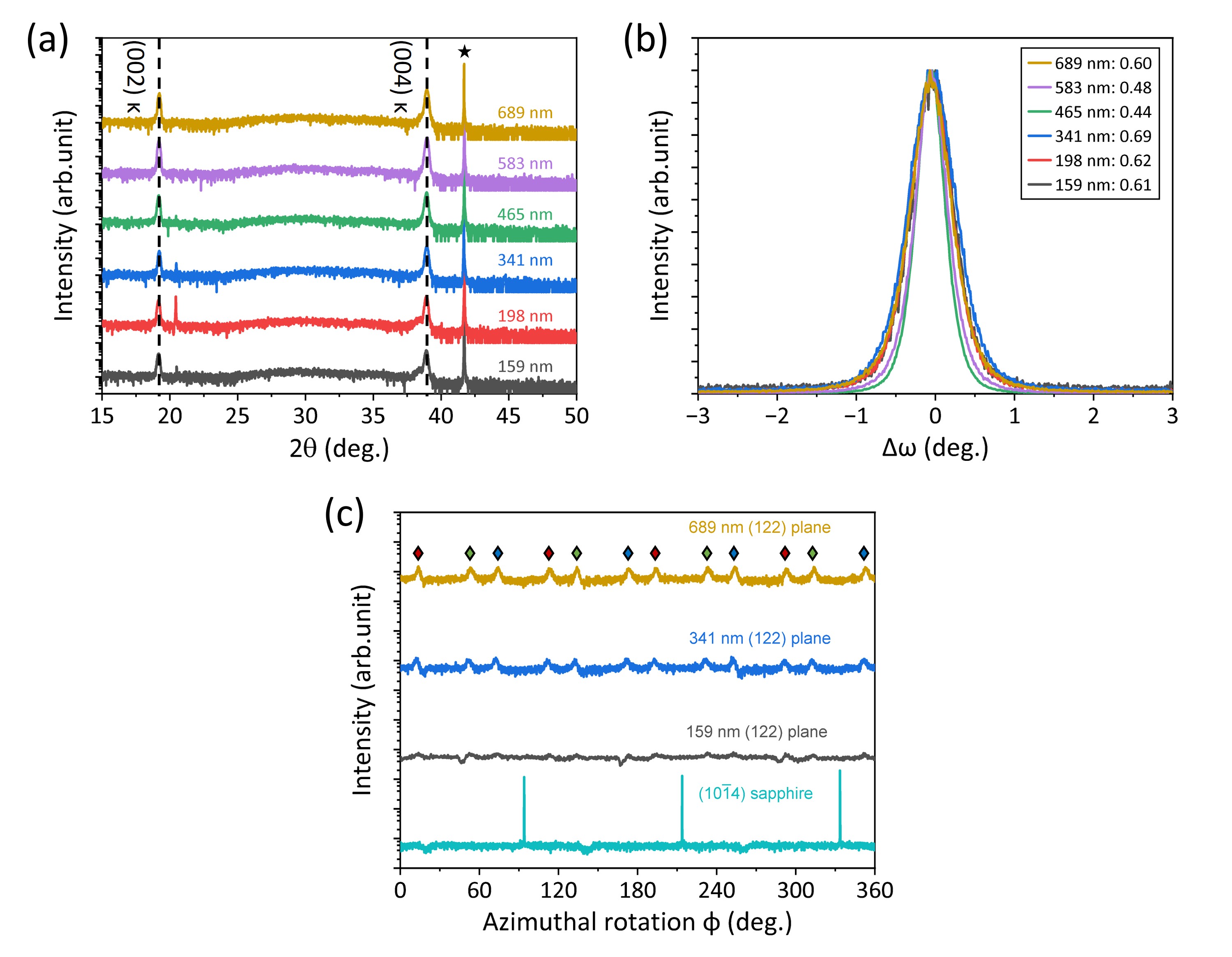}
    \caption{{\color{black}\textbf{(a)} HR-XRD 2$\theta$-$\omega$ scans for the six $\kappa$-Ga\textsubscript{2}O\textsubscript{3} films on $c$-plane sapphire substrates. The black star symbol denotes the diffraction peaks from the substrates. \textbf{(b)} Symmetric X-ray rocking curves of the (004) reflection for the six films. \textbf{(c)} $\phi$-scans of the $\{122\}$ reflection for three representative $\kappa$-Ga\textsubscript{2}O\textsubscript{3} films. The diamond symbols in different colors represent three types of in-plane rotational domains.}}
    \label{fig:XRD}
\end{figure}

High-resolution X-ray diffraction (HR-XRD) patterns of the as-grown $\kappa$ films were acquired using a Panalytical X’pert Pro MPD X-ray diffractometer equipped with a four-crystal Ge (220) monochromator. As shown in Fig.\ref{fig:XRD}(a), only (002) and (004) $\kappa$-Ga\textsubscript{2}O\textsubscript{3} peaks are observed other than the substrate peak. {\color{black} Based on our previous cross-sectional transmission electron microscopy studies on a representative 700 nm thick $\kappa$-Ga\textsubscript{2}O\textsubscript{3} film grown under the same conditions, $\sim$3 monolayers of pseudomorphic $\alpha$-Ga\textsubscript{2}O\textsubscript{3} form on the $c$-plane sapphire substrate, followed by a $\sim$20 nm transition layer containing a mixture of $\beta$ and $\gamma$ phase, with phase-pure $\kappa$ on top of the transition layer. The ultra-thin pseudomorphic $\alpha$ layer and the low volume fraction and poor crystallinity of the transition layer fall below the detection limit of the XRD scans.\cite{jiang_layered_2022} The presence of such a transition layer is commonly observed in MOCVD-grown $\kappa$-Ga\textsubscript{2}O\textsubscript{3} films on $c$-plane sapphire; \cite{zhuo2017beta,hassa2020control,xu2019strain,cora_real_2017,jiang_layered_2022} these films are thus referred as nominally phase-pure. Figure 1(b) shows that the full-width-at-half-maximum of the X-ray rocking curves for these films is comparable, indicating similar crystallinity across various film thicknesses. Figure 1(c) shows the presence of in-plane rotational domains separated by 120$^{\circ}$ across, independent of film thickness. Since these rotational domains are oriented in-plane, they will not affect the out-of-plane thermal conductivity measured in this study. The presence of such in-plane rotational domains is due to the nature of 6-fold symmetry of the $c$-plane sapphire substrate.\cite{tang_kinetic_2026}} The (001) out-of-plane epitaxial relationship between $\kappa$ and $c$-sapphire is commonly observed due to the similarity of the oxygen atomic arrangement between (0001) sapphire and the (001) $\kappa$. Both exhibit a pseudo-hexagonal lattice with a lattice mismatch of 5.7\%. {\color{black} The reciprocal spacing maps (RSM) for the thin (159 nm), medium (341 nm), and thick (689 nm) films (see Supplementary Information, Fig. S1) show that the films are relaxed to similar states across the entire film thickness. The atomic force microscopy (AFM) images of the films are shown in Fig. S2, where the thinner film shows lower roughness than the thicker film, but all within 1.5\% of the film thickness. These additional results confirm that the crystallinity, domain structure, strain relaxation and relative surface roughness remain consistent across these six films. }

The thickness-dependent and temperature-dependent thermal conductivities of $\kappa$-Ga$_2$O$_3$ were measured using frequency domain thermoreflectance (FDTR), a non-contact optical technique that uses continuous-wave pump (515 nm) and probe (532 nm) lasers to measure the thermal properties. The pump laser, which serves as a temporally periodic, spatially Gaussian heat source, is focused on the sample and absorbed by the Au transducer. The thermoreflectance of Au at 532 nm enables the co-aligned probe laser to monitor the periodic temperature response, which has a phase lag relative to the pump laser that depends on the sample's thermal properties. The phase lag between the pump (heat flux) and probe (temperature) lasers was measured for multiple modulation frequencies ranging from 100 kHz to 5 MHz. Details of this measurement procedure are described in our previous work on $\gamma$-Ga$_2$O$_3$.\cite{park_size_2025} Sensitivity analysis was performed to determine the sensitivity of phase lag to each parameter used in the analytical solution. The sensitivity of a parameter to the phase lag measured in FDTR is quantified as

\begin{equation} \label{eq:1}
S_{\phi} = \frac{d \ln (\phi)}{d \ln (s)} = \frac{d\phi}{ds} \times\frac{s}{\phi},
\end{equation}
which is the normalized change in phase per unit change in the parameter. \cite{gundrum_thermal_2005} A high sensitivity to the targeted fitting parameter indicates that it can be accurately measured by FDTR. If FDTR is sensitive to another input parameter, then that parameter needs to be quantified externally with low uncertainty to minimize its unintentional influence on the fitted parameter. 

\begin{figure}[h]
    \centering
    \includegraphics[width=\linewidth]{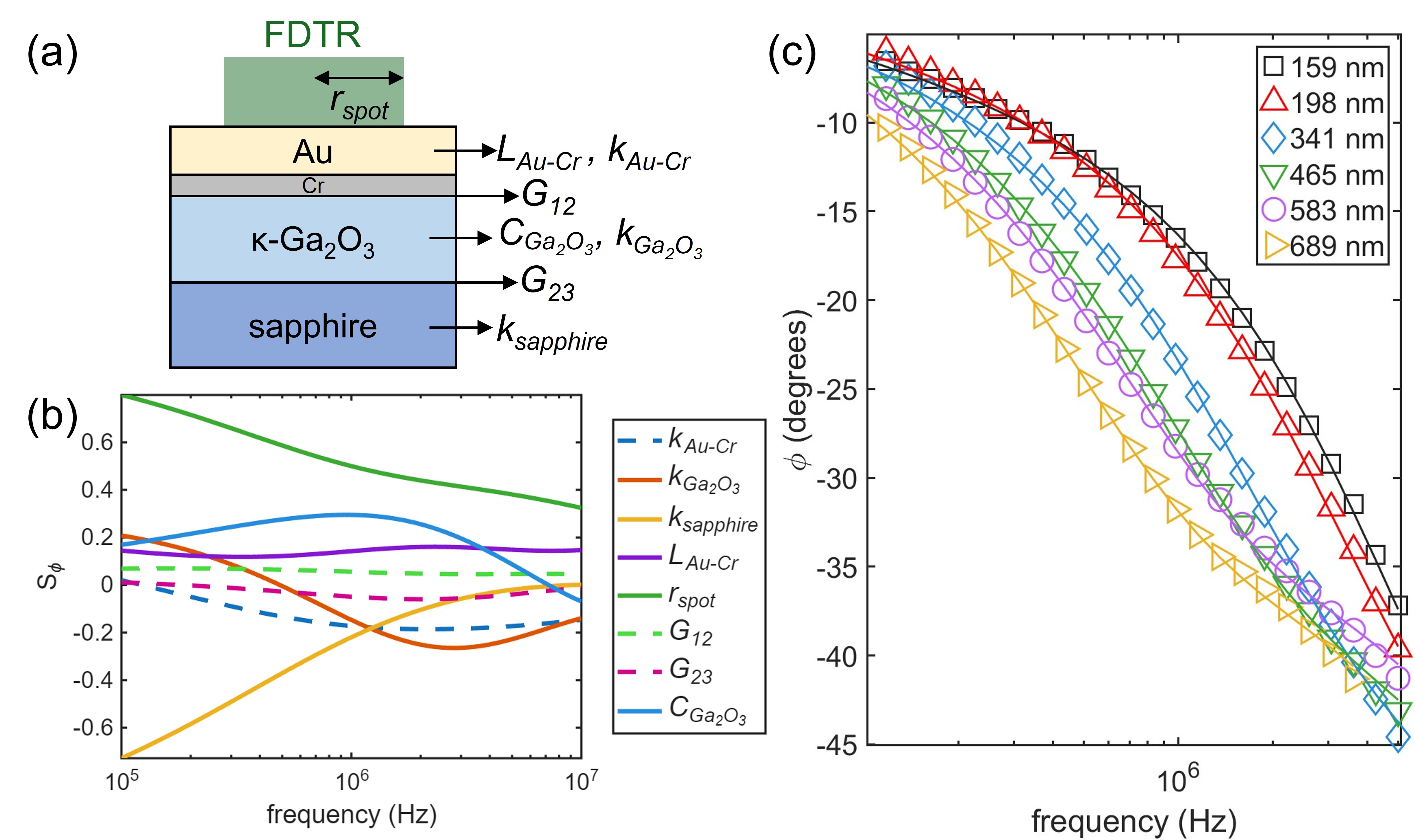}
    \caption{\color{black}\textbf{(a)} Stacking sequence of the sample. \textbf{(b)} Sensitivity analysis for the 159 nm $\kappa$-Ga$_2$O$_3$ film. \textbf{(c)} Measured phase lag and analytical solution fits for thickness-dependent measurements on $\kappa$-Ga$_2$O$_3$ films.}
    \label{fig:sensitivity}
\end{figure}
The sensitivity analysis reveals that the FDTR is sensitive to the volumetric heat capacity of $\kappa$-Ga$_2$O$_3$ ($C_{Ga_2O_3}$), the laser spot radius ($r_{spot}$), and the thermal conductivity of the sapphire substrate ($k_{sapphire}$). $C_{Ga_2O_3}$ ($3090\pm150$ J/cm$^3\cdot$K) was determined from the reference values for the specific heat at $T =$ 293 K ($\sim$506 J/kg$\cdot$K)\cite{ribeiro_da_silva_phonons_2025} and density ($\sim$6.1 g/cm$^3$).\cite{cora_real_2017, playford_structures_2013} $r_{spot}$ ($2.6\pm0.1$ $\mu$m) was measured using a reference sample (Au-Cr transducer on fused silica) prior to the $\kappa$-Ga$_2$O$_3$ measurement. $k_{sapphire}$ (38 $\pm$ 2 W/m$\cdot$K) was measured separately using FDTR and was fixed for the $\kappa$-Ga$_2$O$_3$ measurement. 

A 5 nm adhesion layer of Cr capped with a 65 nm Au transducer layer was deposited on top of the $\kappa$-Ga$_2$O$_3$ films using a Kurt Lesker PVD 75 Electron Beam Evaporator. The true combined thickness of the Au-Cr transducer ($L_{Au-Cr}$) was measured with a KLA Tencor P-15 Profilometer. The composite electrical conductivity of the Au-Cr transducer was measured with the four-point probe (4PP) method. The Wiedemann-Franz law was used to determine the Au-Cr thermal conductivity from the electrical conductivity. Three parameters that cannot be determined externally are $k$\textsubscript{$\kappa$-Ga$_2$O$_3$} and the thermal boundary conductances at the Au-Cr/$\kappa$-Ga$_2$O$_3$ interface ($G_{12}$) and at $\kappa$-Ga$_2$O$_3$/sapphire interface ($G_{23}$). {\color{black}The $\kappa$-Ga$_2$O$_3$ film contains a $\beta$-/$\gamma$-Ga$_2$O$_3$ transition region adjacent to the $c$-sapphire substrate. Cross-sectional TEM reveals substantial variation in the thickness of this region and a diffuse boundary between the transition region and the $\kappa$-Ga$_2$O$_3$  film, without observable voids. Because the transition-layer thickness, composition, and thermal properties cannot be independently characterized, we represent the $\kappa$-Ga$_2$O$_3$ film and transition region as a single effective layer in the thermal model. Therefore, the extracted k represents the effective cross-plane thermal conductivity of the combined $\kappa$-Ga$_2$O$_3$/$\beta$-/$\gamma$-Ga$_2$O$_3$ region, rather than the intrinsic conductivity of phase-pure $\kappa$-Ga$_2$O$_3$.} {\color{black} The sensitivity analysis reveals that the phase is marginally sensitive to $G_{23}$ throughout the entire range of modulation frequencies, as shown in Fig. \ref{fig:sensitivity}(b). 
For this reason, we simultaneously extracted thickness-dependent and temperature-dependent $k$\textsubscript{$\kappa$-Ga$_2$O$_3,eff$} and $G_{12}$, where $k$\textsubscript{$\kappa$-Ga$_2$O$_3,eff$} includes $k$\textsubscript{$\kappa$-Ga$_2$O$_3$} and $G_{23}$. The interfacial conductance $G_{23}$ was estimated by extrapolating the effective thermal resistance of the two thinnest films to zero thickness. We then used this estimated interfacial conductance to separate the interface contribution from the measured effective thermal conductivity and obtain the $\kappa$-Ga$_2$O$_3$ thermal conductivity for each film. Further details are provided in Section IV of the Supplementary Information.} We expect $G_{12}$ to vary over different samples, as this parameter may be influenced by the transducer deposition conditions and the surface roughness.

Uncertainties of the fitting parameters were determined using the Bayesian Parameter Estimation (BPE).\cite{drew_analysis_2026} Previous FDTR uncertainty quantification methods, i.e., 1-variable Root Sum of Squares (RSS) and Monte Carlo,\cite{yang_uncertainty_2016} quantify the uncertainties by finding a single local minimum of the sum of squared error (SSE). BPE, on the other hand, assigns a probability to a given set of parameters based on the SSE of the fit they generate. This probability is further modified by incorporating external information known about the parameters (i.e., the prior distributions). Here we report the median of the resultant probability distributions as the "best-fit" value, and the 5th and 95th percentiles as the lower and upper bounds of the 90$\%$ confidence interval. For our BPE analysis, we combined the Gaussian prior and uniform prior approach. We set two externally determined fixed parameters, $C_{Ga_2O_3}$ ($3090\pm150$ J/cm$^3\cdot$K) and $r_{spot}$ ($2.6\pm0.1$ $\mu$m), as Gaussian priors. We chose these parameters because the phase lag is sensitive to their values. Other control parameters were fixed in the BPE analysis because the phase lag is insensitive to their values and their values are well known. In the Gaussian prior, the weighting factor is equivalent to a Gaussian distribution centered at the externally determined value of our parameters, with standard deviations equal to the uncertainty in each parameter. Then, we set two fitting parameters, $k$\textsubscript{$\kappa$-Ga$_2$O$_3$} and $G_{12}$, as uniform priors. In the uniform prior, all values were considered equally likely, which is more suitable for unknown parameters. {\color{black}Uniform prior bounds were initially centered on preliminary best-fit estimates solely to define a computationally tractable parameter grid; these estimates were not used as constraints in the Bayesian posterior evaluation. We verified prior-bound sensitivity by repeating the BPE analysis with systematically varied ranges and selected final bounds that fully contain the posterior distributions, confirming that the reported thermal conductivities are not driven by the chosen prior limits.}

\begin{figure}[h]
    \centering
    \includegraphics[width=\linewidth]{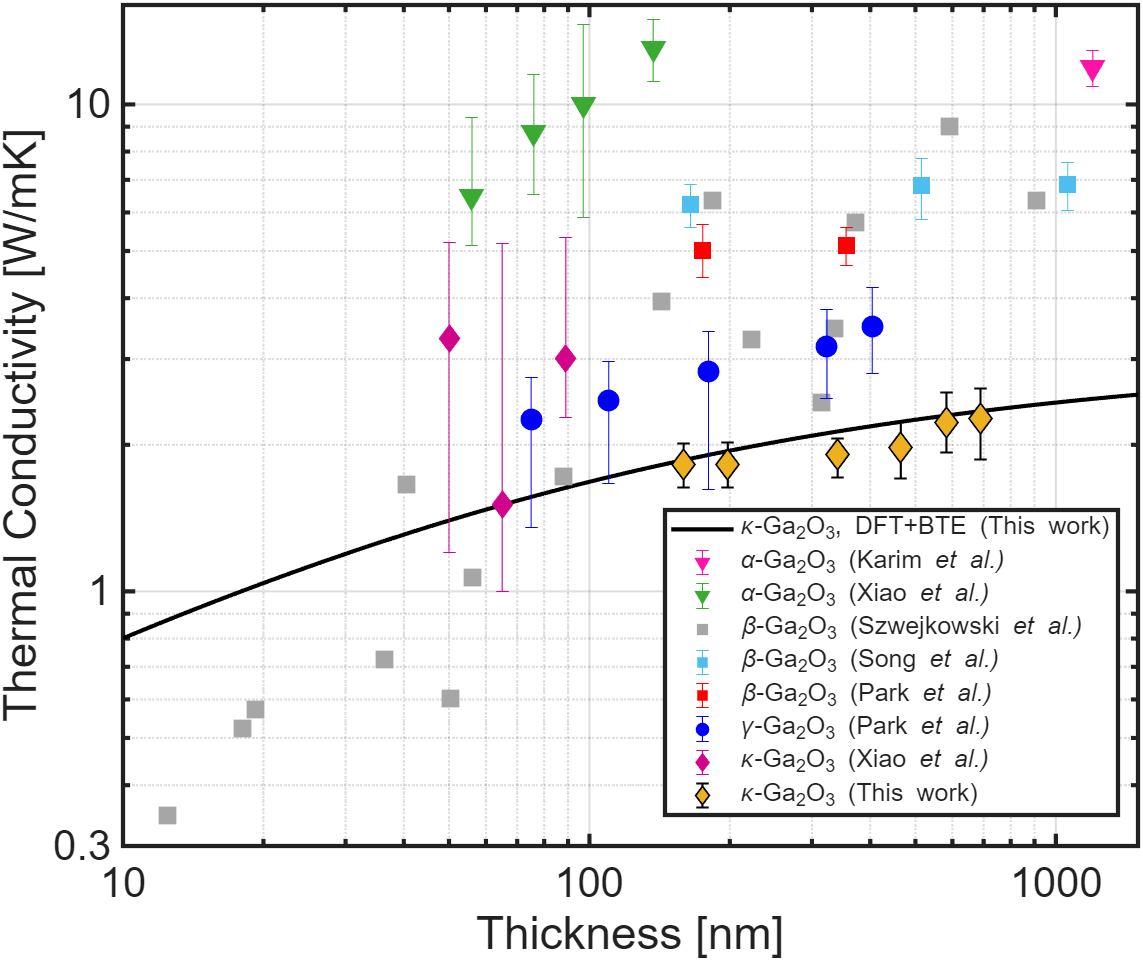}
    \caption{\color{black}Thickness-dependent thermal conductivity of Ga$_2$O$_3$ polymorphs. Inverted triangle symbols are $\alpha$-Ga$_2$O$_3$ from Karim \textit{et al.} (pink)\cite{karim_thermal_2023} and from Xiao \textit{et al.} (green).\cite{xiao_phasedependent_2024}
    Square symbols are $\beta$-Ga$_2$O$_3$ from our previous work (red),\cite{park_size_2025} Szwejkowski \textit{et al.} (gray),\cite{szwejkowski_size_2015} and Song \textit{et al.} (light blue).\cite{song_thermal_2021} Circle symbols are $\gamma$-Ga$_2$O$_3$ from our previous work (blue).\cite{park_size_2025}  Diamond symbols are $\kappa$-Ga$_2$O$_3$ measured in this work (yellow) and from Xiao \textit{et al.} (purple).~\cite{xiao_phasedependent_2024} Our first-principles-based calculations, which use the interatomic force constants of Yang \textit{et al.},~\cite{yang_lattice_2024} are displayed as a black line.} 
    
    \label{fig:L_dependent}
\end{figure}

The thickness-dependent thermal conductivity of $\kappa$-Ga$_2$O$_3$ is compared to that of other polymorphs in Fig. \ref{fig:L_dependent}. {\color{black}
Although the 90\% confidence intervals of the $\kappa$-Ga$_2$O$_3$ films and $\gamma$-Ga$_2$O$_3$ measurements partially overlap, the median cross-plane thermal conductivity values of the $\kappa$-Ga$_2$O$_3$ films are lower than other experimentally measured Ga$_2$O$_3$ polymorphs at comparable thicknesses. 
At comparable film thicknesses, the $\kappa$-Ga$_2$O$_3$ films exhibit lower median values of cross-plane thermal conductivity than the other experimentally measured Ga$_2$O$_3$ polymorphs.} This observation is consistent with lattice thermal conductivities predicted by first principles\cite{yang_lattice_2024} and the only prior measurements.\cite{xiao_phasedependent_2024} 
One outlier is the computational study that reported an extremely low lattice thermal conductivity of $\gamma$-Ga$_2$O$_3$ ($\sim$0.12 W/m$\cdot$K).\cite{sun_influence_2024} The size dependence in the thermal conductivity of $\kappa$-Ga$_2$O$_3$ is present {(\color{black} 1.83 $^{+0.19}_{-0.18}$ W/m$\cdot$K to 2.26 $^{+0.35}_{-0.40}$ W/m$\cdot$K from 159 nm to 689 nm)} but less prominent than $\beta$- and $\gamma$-Ga$_2$O$_3$. 

{\color{black} First-principles-based calculations were performed to predict the size dependence of thermal conductivity in phase-pure $\kappa$-Ga$_2$O$_3$ films using the phonon Boltzmann transport equation. Previous first-principles-based studies have calculated the phonon properties and bulk thermal conductivity of $\kappa$-Ga$_2$O$_3$.\cite{yang_lattice_2024, xiao_phasedependent_2024, sun_influence_2024} Using the second- and third-order interatomic force constants provided to us by Yang et al.,\cite{yang_lattice_2024} we reconstructed the mode-resolved lattice thermal conductivity within the relaxation time approximation. To account for the finite film thickness, phonon-boundary scattering was added to the intrinsic phonon-phonon scattering rate\cite{mcgaughey_size-dependent_2011}  using the Matthiessen rule. The thickness-dependent thermal conductivity was then obtained by recalculating the contribution of each phonon mode using the corrected phonon lifetimes and summing over all modes. Further details about the calculations are provided in Section VIII of the Supplementary Information. As shown in Fig. \ref{fig:L_dependent}, our measured thermal conductivities follow the calculation trend of increasing thermal conductivity with increasing film thickness. Four of the six measured points capture the calculations within their uncertainties. The measured thermal conductivities are, on average, 12\% lower than the calculations. This discrepancy likely results from internal defects and/or the thin transition layer that contains other Ga$_2$O$_3$ phases, which introduce additional phonon scattering.}

The weak size dependence observed in the thermal conductivity of $\kappa$-Ga$_2$O$_3$ is attributed to its intrinsically short phonon MFPs.\cite{yang_lattice_2024, xiao_phasedependent_2024, sun_neuroevolution_2023} First-principles studies have shown that the phonon frequency spectrum of $\kappa$-Ga$_2$O$_3$ is significantly narrower than that of $\beta$-Ga$_2$O$_3$, which results in lower phonon group velocities.\cite{wang_dissimilar_2024} This behavior arises from the complex crystal structure of the $\kappa$ phase. Compared with the $\alpha$ and $\beta$ phases, $\kappa$-Ga$_2$O$_3$ exhibits more distorted Ga-O polyhedra and a broader distribution of Ga-O bond lengths, which lowers the average bond stiffness and reduces phonon group velocities.\cite{yang_lattice_2024} In addition, first-principles calculations report larger third-order force constants and Grüneisen parameters in $\kappa$-Ga$_2$O$_3$, indicating stronger phonon anharmonicity and enhanced phonon-phonon scattering rates.\cite{yang_lattice_2024} The structural complexity also leads to lower phonon participation ratios in $\kappa$-Ga$_2$O$_3$, suggesting that a larger fraction of vibrational modes are spatially localized and contribute weakly to thermal transport.\cite{sun_neuroevolution_2023} 

Since the thermal conductivity of thin-film $\kappa$-Ga$_2$O$_3$ is dominated by phonon-phonon scattering, increasing the film thickness is not expected to enhance its thermal conductivity significantly. This limitation hinders effective heat dissipation in $\kappa$-Ga$_2$O$_3$-integrated devices. Instead, when thermal management is critical, the low thermal conductivity of $\kappa$-Ga$_2$O$_3$ can be complemented by growing it on high thermal conductivity polycrystalline diamond\cite{girolami_structural_2024} or SiC\cite{xu_efficient_2021,song_ga2_2021} substrates. 
 
\begin{figure}[h]
    \centering
    \includegraphics[width=1\linewidth]{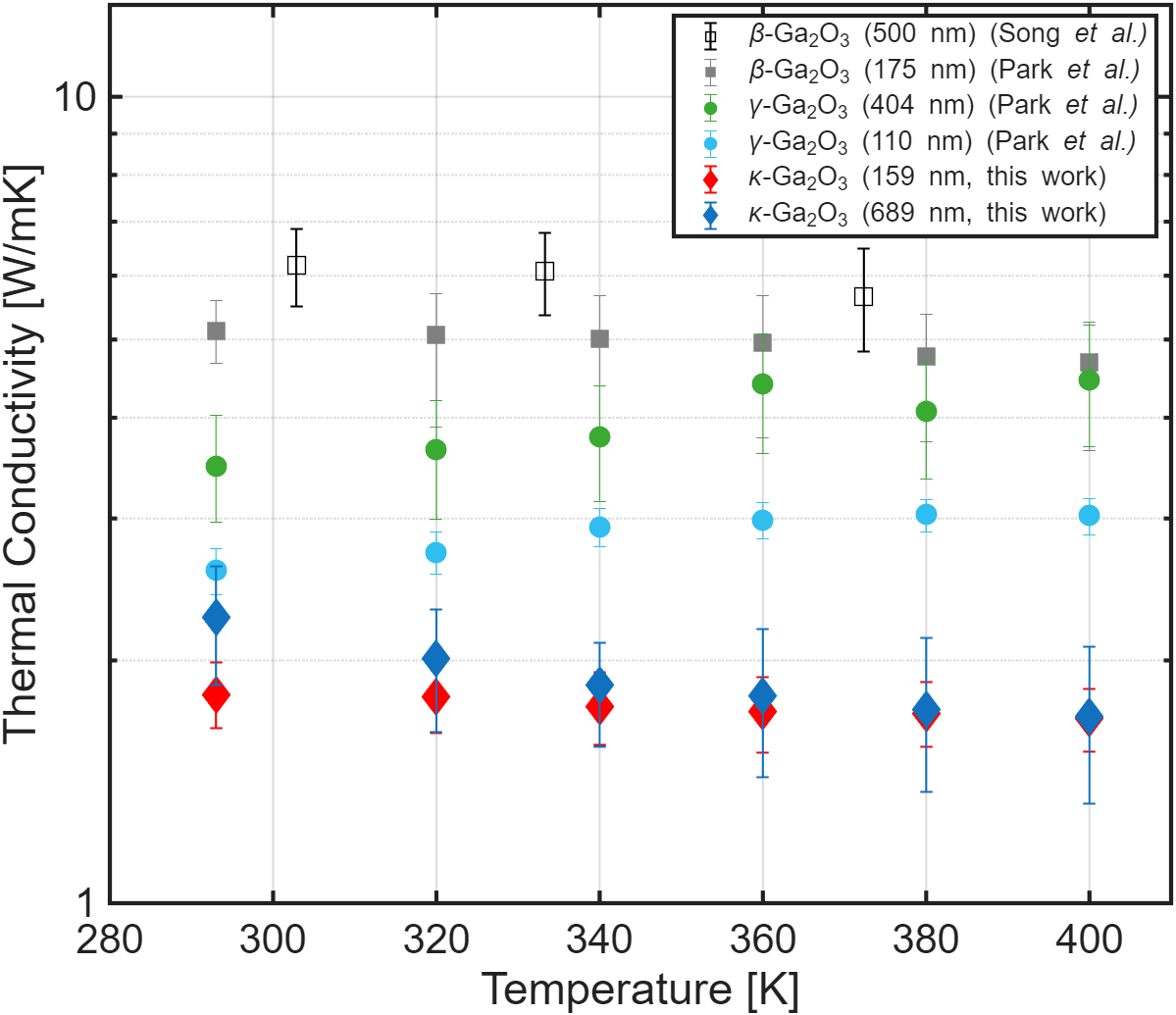}
    \caption{\color{black}Temperature-dependent thermal conductivity of Ga$_2$O$_3$ polymorphs. Square symbols are $\beta$-Ga$_2$O$_3$ from our previous work (gray)\cite{park_size_2025} and Song \textit{et al.} (black empty).\cite{song_thermal_2021} Circle symbols are $\gamma$-Ga$_2$O$_3$ from our previous work.\cite{park_size_2025} Diamond symbols are $\kappa$-Ga$_2$O$_3$ measured in this work.}
    \label{fig:T_dependent}
\end{figure}

The temperature-dependent thermal conductivities of the {\color{black}159 nm and 689 nm} $\kappa$-Ga$_2$O$_3$ films from 293 K to 400 K are shown in Fig. \ref{fig:T_dependent}. For both films, the thermal conductivity decreases with increasing temperature due to enhanced phonon-phonon scattering. The junction temperature of Ga$_2$O$_3$-based devices can reach up to 200 $^\circ$C,\cite{chatterjee_device-level_2019} where the thermal conductivity of $\kappa$-Ga$_2$O$_3$ is further reduced, highlighting a challenge for effective thermal management. {\color{black} The temperature-dependent thermal conductivity data were fitted using $k = AT^{-n}$, as shown in Fig. S11, yielding $n=0.23$ (95$\%$ CI: 0.17-0.29) for the 159 nm film and $n=0.93$ (95$\%$ CI: 0.61-1.26) for the 689 nm film. The temperature dependence in the thermal conductivity is not as pronounced in the 159 nm film,} where phonon-boundary scattering is more prominent. Since boundary scattering is largely temperature-independent, it suppresses the temperature dependence of the overall thermal conductivity. This trend was also observed in $\beta$-Ga$_2$O$_3$ from our previous measurement and the literature.\cite{song_thermal_2021} In contrast to $\beta$- and $\kappa$-Ga$_2$O$_3$, $\gamma$-Ga$_2$O$_3$ exhibits an increase in thermal conductivity with rising temperature, which is attributed to its disordered structure.\cite{park_size_2025} Temperature‑independent phonon-defect scattering dominates in a disordered material, so the effective mean free path is nearly constant, and the thermal conductivity rises primarily through the increase in heat capacity with temperature.\cite{park_size_2025} 

We measured the cross‑plane thermal conductivity of (001) $\kappa$-Ga$_2$O$_3$ thin films on $c$-plane sapphire substrates over thicknesses from {\color{black}159 nm to 689 nm} using FDTR. Across this thickness range, $k$\textsubscript{$\kappa$-Ga$_2$O$_3$} ranges from {\color{black} 1.83 $^{+0.19}_{-0.18}$ W/m$\cdot$K to 2.26 $^{+0.35}_{-0.40}$ W/m$\cdot$K} and shows a weak thickness dependence, indicating that heat transport in $\kappa$-Ga$_2$O$_3$ is predominantly limited by intrinsically short phonon MFPs rather than phonon‑boundary scattering. {\color{black}We also performed first-principles-based calculations to predict the size dependence in the thermal conductivity of $\kappa$-Ga$_2$O$_3$, and the calculated trend agrees well with the experimental results.} Temperature‑dependent measurements between 293 and 400 K reveal a moderate decrease in $k$\textsubscript{$\kappa$-Ga$_2$O$_3$} due to enhanced phonon-phonon scattering, with the thinner film exhibiting a weaker temperature dependence due to increased phonon-boundary scattering. When compared with literature-reported data and first‑principles calculations for $\alpha$‑, $\beta$‑, and $\gamma$-Ga$_2$O$_3$, $\kappa$-Ga$_2$O$_3$ exhibits the lowest thermal conductivity at comparable thicknesses, which we attribute to its highly distorted polyhedral network, reduced group velocities, and strong anharmonicity that lead to short phonon MFPs. Our findings highlight the need for alternative thermal management strategies, such as integration with high thermal conductivity substrates (e.g., diamond or SiC) or engineered composite stacks, for $\kappa$-Ga$_2$O$_3$‑based power and RF electronics.

\section*{Supplementary Material} 
See the supplementary material for thin film strain and surface morphology characterization (Figs. S1 and S2); film thicknesses and associated uncertainties (Table S1); parameters used in the BPE uncertainty analysis (Table S2) and BPE posterior results for all six films (Figs. S3-S8); BPE posterior results for the thickness-dependent measurements (Table S3) and the temperature-dependent measurements of the 159 nm and 689 nm films (Tables S4 and S5, respectively); details of the linear extrapolation method used to estimate $G_{23}$ (Sec. S4 and Fig. S9); thermal conductivity results after removing $G_{23}$ for the thickness- and temperature-dependent measurements (Tables S6 and S7, respectively); phase lag data and analytical solution fits for the temperature-dependent measurements of the 159 nm and 689 nm films (Figs. S10 and S11, respectively); a power-law analysis of the temperature-dependent thermal conductivity (Fig. S12); sensitivity analyses for the 341 nm and 689 nm films (Figs. S13 and S14, respectively); and details of the calculations used to solve the phonon Boltzmann transport equation (Sec. S8).

\section*{Author's Contributions}
\noindent\textbf{Seungwon Park}: Conceptualization (lead); Data curation (lead); Formal analysis (lead); Methodology (lead); Software (lead); Visualization (lead); Validation (lead); Writing - original draft (lead); Writing - review $\&$ editing (lead). \textbf{Jingyu Tang}: Conceptualization (equal); Data curation (equal); Formal analysis (equal); Visualization (equal); Writing - original draft (equal); Writing - review $\&$ editing (equal). {\color{black}\textbf{Minsung An}: Data curation (support); Formal analysis (lead); Methodology (support); Software (support); Validation (support); Writing - review $\&$ editing (support).  \textbf{Alan J. H. McGaughey}: Funding acquisition (lead); Project administration (lead); Supervision (lead); Writing - review $\&$ editing (lead).} \textbf{Lisa M. Porter}: Funding acquisition (lead); Project administration (lead); Supervision (lead); Writing - review $\&$ editing (lead). \textbf{Robert F. Davis}: Funding acquisition (lead); Project administration (lead); Supervision (lead); Writing - review $\&$ editing (lead). \textbf{Jonathan A. Malen}: Funding acquisition (lead); Project administration (lead); Supervision (lead); Writing - review $\&$ editing (lead).

\section*{Acknowledgments}
This work was supported by the Army Research Office Award No. W911NF2220191. The authors acknowledge the use of the Bertucci Nanotechnology Laboratory at Carnegie Mellon University supported by grant BNL-78657879 and use of the Materials Characterization Facility at Carnegie Mellon University supported by grant MCF-677785. \color{black}{The authors thank Jinkyu Kim for performing the AFM imaging and surface morphology characterization. The authors also thank Dr. Jinfeng Yang for providing the first-principles calculation data.}

\section*{Conflict of Interest}
The authors have no conflicts to disclose.

\section*{Data Availability}
The data that support the findings of this study are available from the corresponding author upon reasonable request. 

\bibliography{References_main}
\end{document}